\documentclass[12pt]{article}
\usepackage{xcolor}
\usepackage{graphicx}
\usepackage{float}
\usepackage{amsthm,comment}
\usepackage{dsfont}
\usepackage{url}
\usepackage[ruled,vlined]{algorithm2e}
\usepackage{enumitem}
\RequirePackage{amsthm,amsmath,amsfonts,amssymb}
\RequirePackage[authoryear]{natbib}
\usepackage[ruled,vlined]{algorithm2e}
\usepackage{setspace}
\usepackage{tabularx} 
\usepackage{booktabs}
\usepackage{authblk}
\usepackage{bbm}
\usepackage{caption,subcaption}
\usepackage{footnote}
\usepackage{wrapfig}
\usepackage{mathtools}
\usepackage{lscape}

\newtheorem{proposition}{Proposition}

\newtheorem{theorem}{Theorem}[section]
\newtheorem{lemma}[theorem]{Lemma}
\theoremstyle{remark}

\makeatletter
\newcommand*{\rom}[1]{\expandafter\@slowromancap\romannumeral #1@}
\makeatother

\RequirePackage[colorlinks,citecolor=blue,urlcolor=blue]{hyperref} 
\newcommand{\blind}{1}

\date{}

\begin{document}

\def\spacingset#1{\renewcommand{\baselinestretch}%
{#1}\small\normalsize} \spacingset{1}

\if1\blind
{
	 \begin{center} 
	\spacingset{1.5} 	{\LARGE\bf  Conditional Inference for Secondary Outcomes Based on the Testing Result for the Primary Outcome in Clinical Trials} \\ \bigskip \bigskip
		\spacingset{1} 
		{\large Tianyu Pan$ ^{1,2} $, Vivek Charu $ ^{2,3} $, Ying Lu $^{1}$, Lu Tian$ ^{1*} $} \\ \bigskip
	 {$^1$Department of Biomedical Data Science, Stanford University School of Medicine\\ 
     $ ^2 $Department of Pathology, Stanford University School of Medicine\\
     $^3$ Quantitative Sciences Unit, Department of Medicine, Stanford University School of Medicine\\
    }
	\end{center}
} \fi

\if0\blind
{
  \bigskip
  \bigskip
  \bigskip
  \begin{center}
  \spacingset{1.5}
    {\LARGE\bf Conditional Inference for Secondary Outcomes Based on Primary Outcome Significance in Clinical Trials}
\end{center}
  \medskip
} \fi

\spacingset{1.8} 


\begin{abstract}
In clinical trials, inferences on clinical outcomes are often made conditional on specific \textit{selective processes}. For instance, only when a treatment demonstrates a significant effect on the primary outcome, further analysis is conducted to investigate its efficacy on selected secondary outcomes. Similarly, inferences may also depend on whether a trial is terminated early at interim stage. While conventional approaches primarily aim to control the family-wise error rate through multiplicity adjustments, they do not necessarily ensure the desired statistical property of the inference result, when the analysis is conducted according to a \textit{selective process}.  For example, the conditional coverage level of a regular confidence interval under a \textit{selective processes} can be very different from its nominal level even after adjustment for multiple testing. In this paper, we argue that the validity of the inference procedure conditional on \textit{selective process} is important in many applications. In addition, we propose to construct confidence intervals with correct conditional coverage probability by accounting for related \textit{selective process}. Specifically, our approach involves a pivotal quantity constructed by inversing the cumulative distribution function of a truncated normal distribution induced by the \textit{selective process}. Theoretical justification and comprehensive simulations illustrate the effectiveness of this method in realistic settings. We also apply our method to analyze data from the SPRINT, resulting in more conservative but conditionally valid confidence intervals for the average treatment effect than those originally published.
\end{abstract}
\noindent%
\vfill

\section{Introduction}
\subsection{Review}\label{sec: review}
It is widely acknowledged in clinical trials that the analysis of multiple outcomes, particularly secondary outcomes, should be conditioned on the significance of the analysis result for the primary outcome \citep{o1997secondary}, as the primary outcome serves as the main efficacy indicator for a treatment. While confidence intervals (CIs) for secondary outcomes are often reported, 
their importance 
is frequently downplayed, especially if the primary outcome is nonsignificant. For example, these results are typically deferred to the discussion section per the policies of major medical journals such as the New England Journal of Medicine (NEJM) and the Journal of the American Medical Association (JAMA) (see \citet{bangalore2015everolimus}, \citet{kato2019effect}, \citet{mcmurray2019dapagliflozin}, and \citet{pocock2021primary} for examples). As a consequence, their usage and significance is not entirely clear to the research community other than the vague statement that they should be interpreted cautiously.

Although reported inference results on secondary outcome analyses are ``marginally'' valid, 
it raises concerns about potential bias when they are only used/reported selectively.  For example, if a researcher only care about secondary outcomes when the results for primary outcomes are significant, then these CIs for secondary outcomes may fail to cover true parameters at the nominal level for this researcher. This constitutes a \textit{selective process}, potentially introducing bias in inferences, especially when primary and secondary outcomes are correlated. In an extreme case, if the correlation is close to one, secondary outcomes would almost always appear significant when the primary outcome is significant. A strong correlation is particularly likely when outcomes are defined based on related clinical endpoints, such as the overall survival as the primary endpoint and progression-free survival as a secondary endpoint. This \textit{selective process} can also occur in group sequential designs, where trials can be terminated early at the interim analysis for efficacy or futility, leading to conditional inferences on related parameters. Additional examples could also be explored within our framework when \textit{selective processes} are present.

In the presence of a \textit{selective process}, the inferences on secondary outcomes become inherently conditional. At a broader level, these inferences condition on specific decisions, which are derived from pre-defined clinical endpoints, such as thresholds established for primary outcome estimator. The relationship between these conditional and marginal inferences - the common practices in recent clinical studies - can be expressed as follows,
\begin{equation}\label{Connect}
    \begin{split}
        & \text{pr}(\text{rejection}\mid \theta_2=0) \\
        = &~\text{pr}(\text{rejection}\mid \theta_2=0, \text{positive decision })\times \text{pr}(\text{positive decision}) +\\
        &~\text{pr}(\text{rejection}\mid \theta_2=0, \text{negative decision})\times \text{pr}(\text{negative decision}),
    \end{split}
\end{equation}
where the probability on the left side of the equation corresponds to the marginal type I error under the null that $\theta_2=0$, while the two conditional probabilities on the right side represent the conditional type I errors. These conditional probabilities depend on whether specific decisions, such as “if the primary outcome is significant” or “if the trial was stopped at an interim stage” were met in the context of earlier examples. Although adjusting for multiple testing across endpoints can control the family-wise error rate (FWER) and, by extension, the marginal type I error on the left, the conditional probability $\text{pr}(\text{rejection}\mid \theta_2=0, \text{decision=positive })$ is not inherently controlled and often inflated due to the \textit{selective process}. This is troublesome, when researchers report and use the test result for $\theta_2$ only when the decision from the selective process is positive. Among those cases, actual type I error rate can be substantially higher than the nominal level.  More generally, this problem raises when constructed CIs are treated, interpreted and used differently according to a set of statistics even without an explicit selective process.   We will further provide a concrete examples to illustrate this concern in Section \ref{sec: examples}. This inflation indicates a need for additional adjustments, which form the primary focus of this study.

Inferences associated with conditional probabilities conceptually align with gatekeeping procedures \citep{dmitrienko2003gatekeeping}, where primary and secondary outcomes (or families) are tested sequentially in a pre-determined hierarchy, generally ordered by their statistical power and clinical significance. The initial concept of such sequential testing can be tracked back to \citet{holm1979simple}. Unlike the primary focus in multiple testing \citep{hochberg1987multiple}, where controlling FWER at a desired level through multiplicity adjustments is essential \citep{hochberg1987multiple,bauer1991multiple,koch1996statistical,bauer1998testing,westfall2001optimally}, the hierarchical nature of gatekeeping procedures can automatically maintain FWER control \citep{bretz2009graphical}. Over the past several decades, researchers have explored FWER control within more complex trial designs, such as group sequential designs \citep{glimm2010hierarchical,tamhane2010testing,tamhane2012adaptive1,tamhane2012adaptive2,ye2013group,maurer2013multiple,alosh2014advanced,tamhane2018gatekeeping}. Furthermore, gatekeeping procedures have been studied under stronger error control frameworks, such as controlling the false coverage rate (FCR; \citet{benjamini2005false}). For a thorough review of gatekeeping procedures, readers may refer to \citet{dmitrienko2007gatekeeping,dmitrienko2009gatekeeping,ristl2019methods,proschan2023note} and references therein for extensive discussions. However, while these methods focus on marginal error probabilities, they do not address controlling the conditional error probabilities defined in \eqref{Connect}, nor do they provide procedures for constructing valid CIs for secondary outcomes, conditioned on decisions passed/rejected (e.g., whether primary outcomes were significant or nonsignificant).

To address this question, we introduce a novel approach for constructing CIs for secondary outcomes that ensures exact coverage probability, conditional on specific decisions made based on primary outcome statistics. This conditional CI framework aligns with conditional coverage, a concept previously explored in contexts such as group sequential designs \citep{ohman2003conditional,fan2006conditional}, subgroup analysis using decision tree methods \citep{neufeld2022tree}, and similar scenarios. In this work, we focus primarily on gatekeeping procedures. Our theoretical findings indicate that our approach can be extended to construct conditional CIs with exact coverage probability across various contexts, such as each interim stage in a group sequential trial or each identified subgroup following a data-dependent subgroup discovery, provided the decisions criteria meet certain conditions. 

For the theoretical guarantees, our method assumes that the primary and secondary outcome statistics asymptotically follow a bivariate normal distribution at specific rates, with the covariance estimated using a weakly consistent estimator. Our assumption relaxes the typical normality assumption employed in previous studies, such as \citet{ohman2003conditional,lee2016exact}. The pivotal quantities for constructing secondary outcome CIs are derived from truncated normal distributions, inspired by \citet{lee2016exact}. This construction is based on ordering within the subsample space \citep{siegmund1978estimation} of secondary outcome statistics, conditional on decisions regarding primary outcome statistics. As a result, our method provides valid p-values, as the extremeness of outcomes is well-defined under this framework.

\subsection{Example 1: Post Selection Inference}\label{sec: examples}
As exemplified in \citet{benjamini2005false}, the probability of covering the true parameter among conditional CIs after selection can differ significantly from that among all possible intervals.
To illustrate the impact of a \textit{selective process} on coverage probability, consider the following toy example.

Suppose we have three parameters $\theta_1$, $\theta_2$ and $\theta_3$, with corresponding point estimators $\hat{\theta}_1$, $\hat{\theta}_2$ and $\hat{\theta}_3,$ respectively. For simplicity, we assume that 
$$\left(\begin{array}{c}\hat{\theta}_1\\\hat{\theta}_2 \\ \hat{\theta}_3 \end{array}\right)\sim N\left\{\left(\begin{array}{c}{\theta}_1\\{\theta}_2 \\ {\theta}_3 \end{array}\right), \left(\begin{array}{ccc} 1 & 0.5 & 0.5 \\ 0.5 & 1 & 0.5 \\ 0.5 & 0.5 & 1\end{array}\right)\right\}.$$
We apply a hierarchical testing procedure = a gatekeeping procedure - to test the hypotheses. Specifically, each hypothesis is tested only if the preceding hypothesis is significant, starting with the first hypothesis:
$$H_{01}: \theta_1\le 0, H_{02}: \theta_2\le 0 ~\mbox{and}~ H_{03}: \theta_3\le 0.$$  
By design, this testing procedure controls the one-sided FWER at the 0.05 level. Following common practice, we present the CI for $\theta_i,$ if and only if a test for $H_{0i}: \theta_i\le 0$ has been conducted. Assuming the true parameter values $\theta_1=\theta_2=\theta_3=0$, we generated 1,000,000 sets of $\{\hat{\theta}_1, \hat{\theta}_2, \hat{\theta}_3\}.$  Using the aforementioned testing procedure, we tested $H_{01}$ 1,000,000 times,  $H_{02}$ 50,762 times, and $H_{03}$ 12,442 times. CIs for $\theta_1$, $\theta_2$ and $\theta_3$ were reported based on whether their respective hypotheses were tested.

Without selection, it is known that the one-sided CI, $\hat{\text{I}}_i=(\hat{\theta}_i-1.64, +\infty)$ maintains a 95\% marginal coverage probability for $\theta_i, i=1, 2, 3$, aligning with the left-hand term in \eqref{Connect}. Suppose that the CI, $\hat{\text{I}}_2,$ for $\theta_2$ is reported, only when $H_{01}$ is rejected.  Likewise, the CI, $\hat{\text{I}}_3,$ for $\theta_3$ is reported, only when both $H_{01}$ and $H_{02}$ are rejected.  With this \textit{selective procedure}, the empirical coverage levels of 50,762 reported intervals for $\theta_2$ and 12,442 reported intervals for $\theta_3$ were 75.5\% and 59.5\%, respectively $-$ far below the nominal 95$\%.$
These conditional probabilities correspond to the right-hand term in \eqref{Connect} when the decisions are passed. The results further reveal that the coverage of the conditional CI does not meet the expected level, leading to an inflation of the Type I error conditionally. The FWER control ensures that 
\begin{align*}
0.05& \ge \text{pr}\left( H_{02} \text{ is rejected} \mid \theta_2=0\right)\\
    & = \text{pr}\left(H_{01} \text{ and } H_{02} \text{ are rejected} \right)\\
    & = \text{pr}\left(\theta_2 \notin \hat{\text{I}}_2, H_{01} \text{ is rejected}\right),
\end{align*}
where the last term can be substantially smaller than the conditional non-coverage probability,  $\text{pr}\left( \theta_2 \notin \hat{\text{I}}_2 \mid H_{01} \text{ is rejected}\right),$ which is not controlled by 0.05. In other words, even the stringent FWER control does not ensure the coverage probability of selected CIs. 

In all replications, there were 1,000,000 $\hat{\text{I}}_1,$ 50,762 $\hat{\text{I}}_2,$ and 12,442 $\hat{\text{I}}_3$ reported. The empirical coverage level of all 1,063,204 CIs was 93.6\%. Although this coverage level is not too different from 95\%, it is mainly driven by the correct coverage level of reported $\hat{\text{I}}_1.$ In other words, this should not be interpreted as a proof of the validity of reported CIs for $\theta_2$ and $\theta_3$. 

This example highlights the importance of controlling the conditional coverage probability for reported CIs in the presence of a \textit{selective procedure}. 

\subsection{Example 2: Conditional Inference}
In this section, we present another toy example to demonstrate that even without an explicit selective process, a conditionally valid inference procedure can still be desirable.  Suppose that in a randomized clinical trial, the effect of a new treatment is evaluated at an interim analysis as well as the final analysis.  Without the loss of generality, one may assume that $\hat{\theta}_{I}$ and $\hat{\theta}_{F}$ are estimators for $\theta_0,$ the parameter measuring the average treatment effect, at the interim and final analyses, respectively, and follow a bivariate Gaussian distribution
$$\left(\begin{array}{c}\hat{\theta}_I\\\hat{\theta}_F \end{array}\right)\sim N\left\{\left(\begin{array}{c}{\theta}_0\\{\theta}_0  \end{array}\right), \left(\begin{array}{cc} 1 & 0.5  \\ 0.5 & 0.5 \end{array}\right)\right\}.$$
Suppose that the futility and efficacy stopping boundaries at the interim analysis are 
$ \hat{\theta}_I < 0.5$ and $\hat{\theta}_I>2.5,$ respectively. In other words, the trial would stop early for futility, if $\hat{\theta}_I < 0.5,$ and for efficacy, if $\hat{\theta}_I>2.5.$  The critical value at the final analysis is $2.02/(2)^{1/2}$ to ensure that overall one-sided type I error is 0.025. 
The CI for $\theta_0$ is constructed and reported at the end of the study.  There are four possible scenarios: the study stops at the interim analysis for futility;  the study stops at the interim analysis for efficacy, the study stops at the final analysis as a positive study,  and the study stops at the final analysis as a negative study.  For the first two scenarios, the reported conventional CI for $\theta_0$ is $[\hat{\theta}_I-1.96, \hat{\theta}_I+1.96].$  For the last two scenarios, the reported CI for $\theta_0$ is $[\hat{\theta}_F-1.96/(2)^{1/2}, \hat{\theta}_F+1.96/(2)^{1/2}].$  We simulate 1,000,000 pairs of $\hat{\theta}_I$ and $\hat{\theta}_F$ with $\theta_0=0.75$ to mimic possible trial results. Based on the simulation, the overall coverage level of 1,000,000 reported 95\% CIs is 93.5\%, which is slightly below the nominal level. However, 
\begin{enumerate}
\item there are 401,055 (40.1\%) trials stopped for futility at the interim analysis, and the empirical coverage level of the 95\% CI for $\theta_0$ is 93.7\%;  
\item there are 40,169 (4\%) trials stopped for efficacy at the interim analysis, and the empirical coverage level of the 95\% CI for $\theta_0$ is 37.4\%;
\item there are 428,082 (42.8\%) trials proceeded to the final analysis with a nonsignificant test result, and the empirical coverage level of the 95\% CI for $\theta_0$ is 99.8\%.
\item there are 130,694 (13.1\%) trials proceeded to the final analysis with a significant test result, and the empirical coverage level of the 95\% CI for $\theta_0$ is 89.4\%;  
\end{enumerate}

It is conceivable that researchers may pay different attention to these CIs depending on the trial outcome. For instance, investigators planning a future study to confirm a hypothesized positive treatment effect may more likely use the reported CI when the trial outcome is positive (i.e., scenarios 2 and 4). However, our simulation indicates that in such cases the CIs can be misleading in neglecting with an actual coverage probability of 77.2\%. More importantly, it is impossible to predict in advance which scenario will occur in practice. Our findings suggest that approximately 17.1\% of trials will fall into scenarios 2 or 4, yielding CIs that have lower-than-expected coverage. Although our previous simulation procedure did not employ Siegmund's ordering \citep{siegmund1978estimation} to ensure exact marginal coverage, the bias induced by these \textit{selective processes} persists even after the adjustment (see Section \ref{sec: convUnif}). In general, a CI would be interpreted within the context of the overall trial result, and different researchers may use them differently. For a potential user of the CI, we want to ensure that the coverage level 
$$ \text{pr}(\theta_0\in \hat{\text{I}}\mid \hat{\text{I}} \text{ is used})=\sum_{r} \text{pr}(\theta_0 \in \hat{\text{I}} \mid r)\times \frac{\text{pr}(\hat{\text{I}} \text{ is used}\mid r)\text{pr}(r)}{\text{pr}(\hat{\text{I}} \text{ is used})}$$
is close to its nominal level, where $\hat{\text{I}}$ represents an estimated 95\% confidence interval for $\theta_0$ and $r$ denotes all possible results of the trial such as ``early futility stopping'', which may affect the use of the CI.  Since one can not control how a reported CI is used, i.e., $\text{pr}(\hat{\text{I}} \text{ is used} \mid r),$ the safest strategy is to ensure the conditional coverage probability $ \text{pr}(\theta_0 \in \hat{\text{I}}\mid r)=95\%$ for all $r.$  In summary, if we believe that the CI would be reported or used differently depending on some summary statistics, maintaining the conditional coverage level is important.  

\section{Methodology}\label{sec: method}

Without loss of generality, we assume a secondary outcome is tested and inferred only if a specific decision is met for a primary outcome estimator. Suppose $\theta_p$ and $\theta_s$ are the true parameters associated with the primary and secondary outcomes, respectively, with corresponding regular estimators $\hat{\theta}_{p}$ and $\hat{\theta}_{s}$ depending on sample size $n$. The secondary outcome is tested and inferred only when $\hat{\theta}_{p} > c_{n}$, where $c_{n}$ is a positive sequence. 

First, we consider the method of constructing CI for $\theta_s$ in a simple setting, where $(\hat{\theta}_{p},\hat{\theta}_{s})^T$ follows a bivariate normal distribution with mean $(\theta_p, \theta_s)^T$ and covariance matrix $n^{-1}\Sigma,$ and
\begin{equation}\label{Sigma}
    \begin{split}
        & \Sigma \equiv \begin{pmatrix}
\sigma_p^2 & \sigma_{ps}\\
\sigma_{ps} & \sigma_s^2
\end{pmatrix}
    \end{split}
\end{equation}
is known. Using the decomposition described in Lemma 5.1. of \citet{lee2016exact}, the inference on $\theta_{s}$ can be made based on the conditional distribution
\begin{equation}\label{decomp}
    \begin{split}
 & \hat{\theta}_{s}\biggm| \frac{\sigma_{ps}}{\sigma_s^2}\times \hat{\theta}_{s} > c_{n} - T_n, T_n=t ,\\
    \end{split}
\end{equation} 
where 
$$T_n \coloneqq \hat{\theta}_{p} - \frac{\sigma_{ps}}{\sigma_s^2}\times \hat{\theta}_{s}$$ and $t$ is observed value of $T_n.$
Since $(\hat{\theta}_{p},\hat{\theta}_{s})^T$ is bivariate normal, and $\hat{\theta}_{s}$ and $T_n$ are uncorrelated, and hence independent. Therefore, this conditional distribution is a truncated normal distribution $N(\mu_s, \sigma_s^2)$ with truncation bounds $(a(t),b(t))$ defined as
\begin{equation}\label{abBounds}
    \begin{split}
        & (a(t),b(t)) = 
    \left\{
        \begin {aligned}
             & \left(\frac{c_{n} - t}{\sigma_{ps}/\sigma_s^2}, +\infty\right) \quad & \sigma_{ps} > 0, \\
             & \left(-\infty,\frac{c_{n} - t}{\sigma_{ps}/\sigma_s^2}\right) \quad & \sigma_{ps} < 0,\\
             & \left(-\infty,+\infty\right) \quad & \sigma_{ps} = 0.
        \end{aligned}
    \right.
    \end{split}
\end{equation}

This demonstrates that the decision rule imposed on the primary outcome estimator effectively translates to a truncation on the secondary outcome estimator, inducing a pivotal quantity through the cumulative distribution function (CDF) inversion of a truncated normal distribution. Let $\text{TN}(\mu, \sigma, a, b)$ denote a truncated normal distribution with mean $\mu$, standard deviation $\sigma$, and truncation bounds $a$ and $b$, with CDF denoted by $G_{\mu, \sigma}^{[a, b]}(x)$. 
We apply the inverse CDF method to the conditional variable in \eqref{decomp}, yielding a uniformly distributed random variable,
\begin{equation}\label{UnifSpec}
    \begin{split}
        & G_{\theta_s,\sigma_s}^{[a(t),b(t)]}(\hat{\theta}_{s}) \biggm|  \hat{\theta}_{p} > c_{n}  \sim U(0,1).
    \end{split}
\end{equation}
As shown in Lemma A.1. of \citet{lee2016exact}, $G_{\mu,\sigma}^{[a,b]}(x)$ is monotonically decreasing in $\mu$ for fixed $a,b,\sigma$ and $x$. In other words, we can find $\hat{L}_{\theta_s}$ and $\hat{U}_{\theta_{s}}$ such that 
\begin{align*} 
G_{\hat{L}_{\theta_s},\sigma_s}^{[a(t),b(t)]}(\hat{\theta}_{s}) & = \frac{\alpha}{2}\\ G_{\hat{U}_{\theta_s},\sigma_s}^{[a(t),b(t)]}(\hat{\theta}_{s}) &= 1 - \frac{\alpha}{2},
\end{align*}
we can form an interval $\left[\hat{L}_{\theta_s}, \hat{U}_{\theta_s}\right]$ as an  $1-\alpha$ level CI for $\theta_s$ with a valid conditional coverage level, i.e., 
$$P\left(\theta_s\in \left[\hat{L}_{\theta_s}, \hat{U}_{\theta_s}\right] \mid \hat{\theta}_{p}>c_{n}\right)=1-\alpha.$$

There are two remaining questions to answer in practice: (1) the estimator $(\hat{\theta}_{p},\hat{\theta}_{s})^T$ often follows an approximate, rather than exact, bivariate normal distribution, and (2) the covariance $\Sigma$ is typically unknown and must be estimated. In the next section, we demonstrate that the pivotal quantity in \eqref{UnifSpec} asymptotically follows a Uniform distribution if $(\hat{\theta}_{p},\hat{\theta}_{s})^T$ is asymptotically bivariate normal and $\Sigma$ is replaced by a weakly consistent estimator $\hat{\Sigma}$.  

In practical settings, we assume that both $\hat{\theta}_{p}$ and $\hat{\theta}_{s}$ are consistent estimators of $\theta_p$ and $\theta_s,$ respectively. Let 
$$\mu \coloneqq \begin{pmatrix}
\theta_p \\
\theta_s
\end{pmatrix}~\mbox{ and }~\hat{\mu} \coloneqq \begin{pmatrix}
\hat{\theta}_{p} \\
\hat{\theta}_{s}
\end{pmatrix},$$ 
we also assume that as the sample size $n\rightarrow \infty,$ $n^{1/2}(\hat{\mu} - \mu)$ converges weakly to a mean zero bivariate Gaussian distribution with a variance-covariance matrix $\Sigma,$ whose consistent estimator is 
$$\hat{\Sigma} = \begin{pmatrix}
\hat{\sigma}_{p}^2 & \hat{\sigma}_{ps}\\
\hat{\sigma}_{ps} & \hat{\sigma}_{s}^2
\end{pmatrix}.$$ 
Lastly, we let the cutoff value $c_{n}\coloneqq c_p\hat{\sigma}_{p}n^{-1/2}.$
Due to the stochastic nature of $\hat{\Sigma}$, a slight modification is required for the bounds defined in \eqref{abBounds} to ensure desirable asymptotic properties. Specifically, the modified bounds are defined as follows,
\begin{equation}\label{ModabBounds}
    \begin{split}
    (a_n^{\delta_n}(t),b_n^{\delta_n}(t)) = 
    \left\{
        \begin {aligned}
             & \left(\frac{c_{n}- t}{\hat{\sigma}_{ps}/\hat{\sigma}_{s}^2}, +\infty\right) \quad & \hat{\sigma}_{ps} > \delta_n, \\
             & \left(-\infty,\frac{c_{n} - t}{\hat{\sigma}_{ps}/\hat{\sigma}_{s}^2}\right) \quad & \hat{\sigma}_{ps} < -\delta_n,\\
             & \left(-\infty,+\infty\right) \quad & |\hat{\sigma}_{ps}| \leq \delta_n,
        \end{aligned}
    \right.
    \end{split}
\end{equation}
where $\delta_n = o(1) $ such that $0 < \delta_n < |\sigma_{ps}|$ for sufficiently large $n,$ when $|\sigma_{ps}| > 0$. In Section \ref{sec: simul}, the simulation results indicate that the $\delta_n$ values are not influential on the inference, even taken as 0. This modification ensures the convergence of $(a_n^{\delta_n}(t),b_n^{\delta_n}(t))$ toward $(a(t),b(t))$, as defined in \eqref{abBounds}. With this modification, the $(1-\alpha)$ CI for $\theta_s$ becomes $\left[\hat{L}_{\theta_s}^{\delta_n}, \hat{U}_{\theta_s}^{\delta_n} \right]$
\begin{align*} 
G_{\hat{L}^{\delta_n}_{\theta_s},\hat{\sigma}_{s}}^{[a_n^{\delta_n}(t),b_n^{\delta_n}(t)]}(\hat{\theta}_{s}) & = \frac{\alpha}{2}\\ G_{\hat{U}^{\delta_n}_{\theta_s},\hat{\sigma}_{s}}^{[a_n^{\delta_n}(t),b_n^{\delta_n}(t)]}(\hat{\theta}_{s}) &= 1 - \frac{\alpha}{2}.
\end{align*}

\section{Theoretical results}
In this section, we discuss the theoretical property of the constructed CI. To this end, we denote the CDF of a bivariate normal distribution $\mathcal{N}(\mu,\Sigma)$ by $\Phi_{\mu,\Sigma}(\cdot,\cdot),$
the CDF of $(\hat{\theta}_{p}, \hat{\theta}_{s})^T$ by $F_n(\cdot, \cdot)$, and the conditional CDF of $T_n$ given $\hat{\theta}_{p} > c_{n}\coloneqq c_p\hat{\sigma}_{p}n^{-1/2}$ by
\begin{equation}\label{CondLaw}
    \begin{split}
        & H_{T_n\mid \hat{\theta}_{p} > c_{n}}(t) = \frac{\text{pr}(T_n\leq t, \hat{\theta}_{p} > c_{n})}{\text{pr}(\hat{\theta}_{p} > c_{n})},
    \end{split}
\end{equation}
where $c_p$ is a pre-fixed positive constant. The following conditions are considered for our theoretical results,
\begin{enumerate}
    \item[(C1)] The cumulative distribution function (CDF) of $\hat{\mu}$, $F_n(\cdot,\cdot)$, satisfies a uniform convergence toward $\Phi_{\mu,n^{-1}\Sigma}(\cdot,\cdot)$ at a specific rate, namely,
    \begin{equation}\label{CDFConverge}
        \begin{split}
            & \sup_{(x_p, x_s)\in R^2}\left|F_n(x_p,x_s) - \Phi_{\mu,n^{-1}\Sigma}(x_p,x_s)\right| \leq \frac{C^*}{n^{\gamma }},
        \end{split}
    \end{equation}
    where $C^*$ and $\gamma$ are positive constants.
    \item[(C2)] $\hat{\Sigma}_n$ is a weakly consistent estimator of $\Sigma$ and $\min(\sigma_s, \sigma_p)>0.$
    \item[(C3)] The probability of observing $\hat{\theta}_{p} > c_{n}\coloneqq c_p\hat{\sigma}_{p}n^{-1/2}$ is at least $\alpha_p$, with $\alpha_p > 0$.
\end{enumerate}

We proceed by interpreting conditions (C1) - (C3).  First, (C2) is a mild condition to ensure that $\hat{\theta}_p$ and $\hat{\theta}_s$ are ``regular'' estimators.  Condition (C3) holds for any fixed $c_0,$ if $\theta_0\ge 0.$ In condition (C1), it is possible to select a sufficiently small $\gamma$. This implies condition (C1) can be verified using Edgeworth expansion, the same technique for showing the validity of bootstrapping estimators, in various settings including linear regression \citep{navidi1986edgeworth}, logistic regression \citep{das2020second}, Cox regression \citep{gu1992edgeworth}, etc. In the following, we present two Propositions to show that some commonly used estimators for average treatment effect satisfy this condition. The proof is provided in the Supplementary Material. 

\begin{proposition}\label{prop0}
    Suppose that $(Y_{i1}, Y_{i2}, R_i)_{i=1}^{n}$ are $n$ $i.i.d.$ observations, where $Y_i=(Y_{i1}, Y_{i2})^T$ is a pair of outcomes with a finite 3rd moment and  $R_i$ is a treatment indicator following Bernoulli distribution with $P(R_i=1)=p_0.$ Let $\theta_{p}=E(Y_{i1}|R_i=1)-E(Y_{i1}|R_i=0)$ and $\theta_s=E(Y_{i2}|R_i=1)-E(Y_{i1}|R_i=0)$ are two parameters measuring the average treatment effect on $Y_{i1}$ and $Y_{i2},$ respectively. $\theta_p$ and $\theta_s$ can be estimated by 
    $$\hat{\theta}_{p}=\frac{1}{n_1}\sum_{i=1}^n Y_{i1}R_i -\frac{1}{n_0} \sum_{i=1}^n Y_{i1}(1-R_i) ~\mbox{ and }~ \hat{\theta}_{s}=\frac{1}{n_1}\sum_{i=1}^n Y_{i2}R_i -\frac{1}{n_0} \sum_{i=1}^n Y_{i2}(1-R_i),$$
    respectively, where $n_1=\sum_{i=1}^n R_i$ and $n_0=n-n_1.$  Then condition (C1) is satisfied for $\gamma = \frac{1}{2}.$
\end{proposition}

\begin{proposition}\label{prop1}
    Suppose that $(T_{i1}, T_{i2}, C_i, R_i)_{i=1}^{n}$ are $n$ $i.i.d.$ random vectors, where $T_i=(T_{i1}, T_{i2})^T$ is a pair of time to event outcomes with a finite bound, $C_i$ is a bounded censoring time,  and $R_i$ is treatment indicator following  Bernoulli distribution with $P(R_i=1)=p_0.$   The observed data are 
    $(X_{i1}, \delta_{i1}, X_{i2}, \delta_{i2}, R_i)_{i=1}^n,$ where $X_{ij}=\min(T_{ij}, C_{ij})$
    and $\delta_{ij}=I(T_{ij}<C_{ij}), j=1, 2.$ First, we assume that $T_i$ and $C_i$ are independent conditional on $R_i.$  We also assume that $T_{ij}\mid R_i$ follows the proportional hazards assumption:
    $$ h_1(t|R=1)=h_1(t|R=0)\exp(\theta_p) ~~\mbox{and}~~ h_2(t|R=1)=h_2(t|R=0)\exp(\theta_s),$$
    where $h_j(t|R=r)$ is the hazard function of $T_{ij}\mid R_i=r, r\in \{0, 1\}.$
    Let $\hat{\theta}_{p}$ and $\hat{\theta}_{s}$ be the estimated regression coefficient of $R_i$ in respective Cox regression model. Then condition (C1) is satisfied for $\gamma = \frac{1}{3}.$
\end{proposition}

We present the theoretical results by limiting our scope to the situation where the null hypothesis is $H_0: \theta_p = 0$ with a one-sided alternative $H_a: \theta_p > 0$.  We show in the following theorem that the constructed pivotal quantity follows $U(0,1)$.

\begin{theorem}\label{mainTheorem}
    Under conditions (C1) - (C3), we have
    \begin{equation}\label{UnifRV}
\lim_{n\rightarrow \infty} \int\text{pr}\left(G_{\theta_{s},\hat{\sigma}_{s}}^{[a_n^{\delta_n}(t),b_n^{\delta_n}(t)]}(\hat{\theta}_{s})\leq u \biggm |  \hat{\theta}_{p} > c_{n}\right) d H_{T_n\mid \hat{\theta}_{p} > c_{n}}(t) \stackrel{p}{\to} u 
    \end{equation}
 for $\forall u \in [0,1].$ 
In other words, the distribution of $G_{\theta_s,\hat{\sigma}_{s}}^{[a_n^{\delta_n}(T_n),b_n^{\delta_n}(T_n)]}(\hat{\theta}_{s})$ conditioning on $\hat{\theta}_{p} > c_{n}\coloneqq c_p\hat{\sigma}_{p}n^{-1/2}$ asymptotically follows $\text{U}(0,1).$ 
\end{theorem}

Theorem \ref{mainTheorem} indicates that it is possible to construct an asymptotic pivotal quantity using a truncated normal distribution with a weakly consistent covariance estimator as the plug-in value. A slightly stronger result implies that 
\begin{equation}\label{condStats}
    \begin{split}
        & G_{\theta_s,\hat{\sigma}_{s}}^{[a_n^{\delta_n}(T_n),b_n^{\delta_n}(T_n)]}(\hat{\theta}_{s})\mid \hat{\theta}_{p} > c_{n},T_n
    \end{split}
\end{equation}
follows $U(0,1)$ with an appropriate choice of $\delta_n=o(1)$, for all $T_n$ but a null set under $H_{T_n\mid \hat{\theta}_{p} > c_{n}}(\cdot)$. Our empirical results in Section \ref{sec: simul} indicate that a specific choice of $\delta_n$ is not influential on the distribution simply let $\delta_n = 0$ in practice. 

\section{Simulation Study}\label{sec: simul}

\subsection{Simulation Design}\label{sec: convUnif}
In this section, we investigate the operating characteristics of the proposed procedure with a finite sample. To this end, four simulation settings are designed to mimic potential real-world scenarios and serve three primary purposes: (1) to validate Theorem \ref{mainTheorem} by examining if the statistic is indeed uniformly distributed as the sample size increases, (2) to compare the proposed method with the conventional CI in terms of interval length and coverage level, and  (3) to assess whether conditional inference on $\theta_s$ is sensitive to varying $\delta_n$ values. In these simulations, unless otherwise specified, we set the true values of $\theta_p$ and  $\theta_s$ to zero.  The data generating processes are detailed as follows.

\begin{enumerate}
    \item[] \textbf{Setting 1} The simulated observations are $(Y_{ip},Y_{is})_{i=1}^n,$ where 
    $$\left(\begin{array}{c} Y_{ip} \\ Y_{is}\end{array}\right)=\left(\begin{array}{c} X_{i1} \\ X_{i2}\end{array}\right) \mbox{ and } \left(\begin{array}{cc} 1 & 0\\ 0.500 & 0.146 \end{array}\right)\left(\begin{array}{c} X_{i1} \\ X_{i2}\end{array}\right)$$
     to induce a targeted Pearson's correlation coefficient of $\rho=0$ and 0.4 between $Y_{ip}$ and $Y_{is},$  respectively, and $X_{i1}$ and $X_{i2}$ are generated from independent $t_4$ distribution.   $\hat{\theta}_{p}$ and $\hat{\theta}_{s}$ are the sample averages of $(Y_{ip})_{i=1}^n$ and $(Y_{is})_{i=1}^n$, respectively, for estimating $\theta_p=E(Y_{ip})$ and $\theta_s=E(Y_{is}).$ 
    \item[] \textbf{Setting 2} 
     Let $T_{i1}$ be time to disease progression and $\text{OS}_i=T_{i2}$ be time to death. The progression free survival time (PFS) is defined as $\text{PFS}_i=\min(T_{i1}, T_{i2}).$  $T_{i1}$ and $T_{i2}$ are generated from exponential distribution with the rate parameter $\lambda_{i1}=1 + \lambda_i$ and $\lambda_{i2}=0.5 + \lambda_i$, respectively, where $\lambda_i \sim U(0,0.5)$, represents a subject-specific random effect. In addition, a censoring time $C_i$ is generated from $U(0, 1),$ and a treatment indicator $R_i$ is generated from a Bernoulli distribution with $\text{pr}(R_i=1)=0.5$. Lastly, the simulated observations are $n$ copies of potentially right censored PFS and OS and an treatment indicator:
    $$\left(Y_{i1}=\min(\text{PFS}_i, C_i), \delta_{i1}=I(\text{PFS}_i<C_i), Y_{i2}=\min(\text{OS}_i, C_i), \delta_{i2}=I(\text{OS}_i<C_i), R_i \right)_{i=1}^n. $$ 
    $\hat{\theta}_{p}$ and $\hat{\theta}_{s}$ are estimated log-transformed hazard ratios by regressing $(Y_{i2}, \delta_{i2})$ and $(Y_{i1}, \delta_{i1})$ on $R_i$, respectively. 
    
    \item[] \textbf{Setting 3} The simulated observations are $(Y_{i1},Y_{i2},Y_{i3})_{i=1}^n$, where 
    $$ \left(\begin{array}{c} Y_{i1} \\ Y_{i2} \\ Y_{i3}\end{array}\right)=\left(\begin{array}{ccc} 1 & 0 & 0\\ 0.500 & 0.867 & 0 \\ 0.500 & 0.289 & 0.816 \end{array}\right)\left(\begin{array}{c} X_{i1} \\ X_{i2} \\ X_{i3}\end{array}\right)+ \left(\begin{array}{c} \theta_1\\ \theta_2 \\ \theta_3 \end{array}\right) $$
 where $X_{i1}, X_{i2}$ and $X_{i3}$ are generated from independent $t_4$ distribution. The Pearson correlation between each pair of outcomes is 0.5.  $\hat{\theta}_{j},$ the estimator for $\theta_j=E(Y_{ij}),$ is sample average of $\{Y_{ij}\}_{i=1}^n,$ where $j=1, 2, 3.$ We consider two sets of true parameter values
   $$  \left(\begin{array}{c} \theta_1\\ \theta_2 \\ \theta_3 \end{array}\right) =  \left(\begin{array}{c} 0 \\ 0 \\ 0 \end{array}\right) \text{ or }  \left(\begin{array}{c} 3(2)^{1/2}/10\\ (2)^{1/2}/10 \\ (2)^{1/2}/10 \end{array}\right) $$
     to examine the conditional type I error and power. 
    \item[] \textbf{Setting 4} The first batch of simulated observations $(Y_{ip},Y_{is})_{i=1}^{100}$ are generated as in setting 1,  where  
    $$\left(\begin{array}{c} Y_{ip} \\ Y_{is}\end{array}\right)=\left(\begin{array}{cc} 1 & 0\\ 0.500 & 0.866\end{array}\right)\left(\begin{array}{c} X_{i1} \\ X_{i2}\end{array}\right)+ \left(\begin{array}{c} \theta_{p} \\ \theta_{ s}\end{array}\right),$$
    
    If the initial $\hat{\theta}_p$ does not exceed a rejection boundary \citep{o1979multiple}, a second batch of additional 100 observations $(Y_{ip}, Y_{is})_{i=101}^{200}$ are generated. The rejection boundary is defined as:
    $$\hat{\theta}_{p1} > \frac{2^{1/2}}{10}c_0\hat{\sigma}_{p1},~~\text{where}~c_0~\text{satisfies}~\text{pr}(Z_1> 2^{1/2}c_0, Z_2> c_0) = 0.05,$$ 
    where $\hat{\theta}_{p1}$ and $\hat{\sigma}_{p1}^2$ are the average and the empirical variance of $(Y_{ip})_{i=1}^{100}$, respectively, and $(Z_1, Z_2)^T$ follows bivariate normal distribution with mean zero, unit variance and a covariance of $2^{-1/2}.$ In addition, $\hat{\theta}_{p2}$ and $\hat{\sigma}_{p2}^2$ are the empirical average and variance of $(Y_{ip})_{i=1}^{200}$, respectively. The null hypothesis $\theta_p=0$ is rejected if either $\hat{\theta}_{p1}$ exceeds the rejection boundary at the interim analysis or $\hat{\theta}_{p2}\ge c_0\hat{\sigma}_{p2}/(10(2)^{1/2})$ at the final analysis. Likewise $\hat{\theta}_{sj}$ and $\hat{\sigma}_{sj}$ are defined based on $Y_{is}$ for $j=1, 2.$ We consider two sets of true parameter values
   $$  \left(\begin{array}{c} \theta_p\\ \theta_s \end{array}\right) =  \left(\begin{array}{c} 0 \\ 0 \end{array}\right) \text{ or }  \left(\begin{array}{c} 2^{1/2}/10 \\ 2^{1/2}/30  \end{array}\right) $$
     to examine the conditional type I error and power. 
\end{enumerate}

For the first two settings, we generate $1,000$ pairs of $\hat{\theta}_{s}$  and $\hat{\theta}_{p}$ with $|\hat{\theta}_{p}| > 1.96\hat{\sigma}_{p}n^{-1/2}$ as the selective criterion and construct both the conventional Wald-type and proposed 95\% CIs for $\theta_s$. In settings 1 and 2, ${\sigma}_{ps}$ is estimated analytically via bootstrap method with 10,000 iterations in setting 3. To compare the proposed method with conventional CIs, we calculate, for each method, the empirical coverage probability and the median width of CIs across the 1,000 Monte Carlo replications. Additionally, we assess the approximation of the distribution of the pivotal statistic in Theorem \ref{mainTheorem} to $\text{Unif}(0, 1)$ using a $\ell_1$ distance, defined as
\begin{equation}\label{empMaxNorm}
    \begin{split}
        & \hat{\ell}_{1} = \frac{1}{M}\sum_{l = 1}^M\left|u_l - \hat{\text{p}}\left\{G_{\theta_{s},\hat{\sigma}_{s}}^{[a_n^{\delta_n}(T_n),b_n^{\delta_n}(T_n)]}(\hat{\theta}_{sl}) \leq u_l\right\}\right|,
    \end{split}
\end{equation}
where  $\hat{\text{p}}\left\{G_{\theta_{s},\hat{\sigma}_{s}}^{[a_n^{\delta_n}(T_n),b_n^{\delta_n}(T_n)]}(\hat{\theta}_{sl}) \leq u\right\}$ is the empirical CDF based on the 1,000 generated pairs of $\hat{\theta}_{s}$ and $\hat{\theta}_{p}$, and $\{u_l=(l-1)/(M-1)\}_{l=1}^M$ is a sequence of $M=10,000$ evenly spaced points between $0$ and $1.$ 

For setting 3, we generate $10^6$ sets of $\{\hat{\theta}_{1}, \hat{\theta}_{2}, \hat{\theta}_{3}\}$ and
construct both Wald-type and proposed CIs. We calculate the coverage probabilities and median width of resulting CIs for $\theta_j, j=1, 2, 3.$ For conditional coverage probabilities, we examine
\begin{equation}\label{condProb4}
    \begin{split}
        & \text{pr}(\theta_j \in \hat{\text{I}}_j\mid \text{D}_1, \ldots, \text{D}_{j-1})
    \end{split}
\end{equation}
where $\hat{\text{I}}_j$ denotes respective 95\% confidence intervals for $j\in \{1,2,3\}$ and $\text{D}_k$ refers to the test result for $\theta_{k}$, which takes value S if the null hypothesis $\theta_k=0$ is rejected at the two sided significance level of 0.05, i.e., $|\hat{\theta}_j|\ge 1.96\hat{\sigma}_jn^{-1/2},$ and ``N'' otherwise.  For example, we calculate and report the conditional probability $\text{pr}(\theta_3 \in \hat{\text{I}}_3 \mid D_1=\text{S}, D_2=\text{S}).$ 
Under the alternative, we also estimate the probability that the CI yielded by each method does not cover 0, serving as an estimation of statistical power, which may also be conditional on different scenarios. 

For setting 4, we generate $10^6$ sets of $\{\hat{\theta}_{p1}, \hat{\theta}_{p2}, \hat{\theta}_{s1}, \hat{\theta}_{s2}\}$ and construct one-sided 95\% CIs for $\theta_p$ and $\theta_s$ using proposed as well as conventional methods. Specifically, the conventional Wald type CI of $\theta_p$ is 
$$\left[\hat{\theta}_{p1}-\frac{2^{1/2}}{10}c_0\hat{\sigma}_{p1}, \hat{\theta}_{p1}+\frac{2^{1/2}}{10}c_0\hat{\sigma}_{p1} \right]$$ 
if $\theta_p=0$ is rejected at the interim analysis, and 
$$\left[\hat{\theta}_{p2}-\frac{1}{10(2)^{1/2}}c_0\hat{\sigma}_{p2}, \hat{\theta}_{p2}+\frac{1}{10(2)^{1/2}}c_0\hat{\sigma}_{p2} \right]$$
if $\theta_p=0$ is not rejected at the interim analysis to be consistent with the result of sequential testing. The Wald-type CI for $\theta_s$ is constructed similarly. For this setting, we also calculate another competitor, that is, the 95\% CIs for $\theta_p$, constructed using ordering of the test results \citep{siegmund1978estimation, rosner1988exact, lai2006confidence}. The resulting CI is denoted by ``sequential CI" in Table \ref{tab:5}. We examine conditional coverage probabilities
\begin{equation}\label{condProb5}
    \begin{split}
        & \text{pr}(\theta_p \in \hat{\text{I}}_p\mid \text{D}_I)~\text{and}~\text{pr}(\theta_s \in \hat{\text{I}}_s\mid \text{D}_I,\text{D}_F),
    \end{split}
\end{equation}
where $\hat{\text{I}}_p$ and $\hat{\text{I}}_s$ denote the respective 95\% confidence intervals for $\theta_p$ and $\theta_s$, $\text{D}_I$ takes ``S'', if 
$\hat{\theta}_{p1} > 2^{1/2}c_0\hat{\sigma}_{p1}/10,$
and ``N'' otherwise. Likewise, $\text{D}_F$ take ``S" if $\hat{\theta}_{p2} > c_0 \hat{\sigma}_{p2}/(10(2)^{1/2}),$ and ``N'' otherwise. As in setting 4, we estimate the conditional statistical power under alternatives. Since only one-sided CI is constructed, we report the average lower bound of the CIs to measure the its precision.

\subsection{Simulation Result}

The results for settings 1 are presented in Tables \ref{tab:1}. 
When the primary and secondary outcome estimators are uncorrelated ($\sigma_{ps} = 0$), the conventional Wald CI maintains the $95\%$ coverage probability for $\theta_s$. This result is expected as $\hat{\theta}_{s}$ is asymptotically independent of $\hat{\theta}_{p}$, and thus, the random decision rule $|\hat{\theta}_{p}| > 1.96\hat{\sigma}_{p}n^{-1/2}$, which is based solely on $\hat{\theta}_{p}$, does not affect the inference on $\theta_s$. As a result, the coverage probability for $\theta_s$ conditional on this random decision is the same as that without conditioning. On the other hand, when the primary and secondary outcome estimators exhibit non-zero covariance ($\sigma_{ps} \neq 0$), the conventional Wald-type CI demonstrates under-coverage, evidenced by an approximately greater than $10\%$ decrease in coverage probability in Tables \ref{tab:1}.

The results for setting 2 is reported in Table \ref{tab:3}, which suggests that using bootstrap estimator for the covariance does not compromise the conditional coverage probability of the proposed method. In contrast, the conventional interval fails to maintain the appropriate conditional coverage level, as these two estimators are correlated with a correlation coefficient of 0.568.

The low values of $\hat{\ell}_{1}$ in Tables \ref{tab:1} and \ref{tab:3} suggest that the empirical CDF of the pivotal statistic in Theorem \ref{mainTheorem} approximates the CDF of $\text{Unif}(0,1)$ fairly well, even for a moderate sample size like $n=100$, further confirming the finite sample validity of the proposed method.

While our method ensures correct conditional coverage probability, regardless of the correlation between primary and secondary estimators, this advantage comes with a cost of a wider CI. Specifically, the median interval width often doubles or even triples compared to conventional intervals and its large interquartile range suggests that CIs can be exceedingly wide in some cases. This phenomenon aligns with both empirical \citep{lee2016exact} and theoretical \citep{kivaranovic2021length} findings that a conditional CI can tend toward infinity when the decision boundary is one-sided and the primary statistic is close to the decision boundary. 

In settings 1 and 2, the tuning parameter $\delta_n$ appears to have a minimal impact on the performance of the proposed method, when chosen sufficiently small, which implies that $\delta_n = 0$ can be used without obvious downside. 

In setting 3,  the simulation results with $\theta_j=0, j=1, 2, 3$ are reported in Table \ref{tab:4}. Both the Wald-type CI and our approach achieve their nominal coverage level unconditionally. However, our approach also ensures the correct conditional coverage level in the presence of \textit{selective processes} (i.e., $\text{D}_1,\text{D}_2 \neq \emptyset$). As discussed in Section \ref{sec: convUnif}, the proposed CIs are also wider than their conventional counterpart. However, it is worth noting that under the most common conditions (i.e., scenarios with the highest frequency), the width inflation is less severe. The hierarchical testing procedure maintains FWER at the $\alpha$ level \citep{bretz2009graphical} by testing subsequent hypotheses at $\alpha$ level only if the current hypothesis is rejected at $\alpha$ level. However, Table \ref{tab:4} indicates that this approach does not control Type I error conditionally, with the coverage probability for $\theta_2$ conditional on $\text{D}_1 = \text{S}$ and  $ \theta_3$ conditional $\text{D}_1 = \text{S}, \text{D}_2 = \text{S}$ below 95\%, when  $\theta_1 = \theta_2 = \theta_3 = 0$ aligning the Type I error with the coverage probability of associated CI. Conversely, our approach controls Type I error conditionally, and thus, also ensuring FWER at $\alpha = 0.05$. We also set $\theta_1 =  3(2)^{1/2}/10$ and $\theta_2 = \theta_3 = (2)^{1/2}/10$ in the simulation, ensuring approximately 85\% power for rejecting the null hypothesis $H_0: \theta_1=0.$   In a rare combination ($\text{D}_1 = \text{N},~\text{D}_2 = \text{S}$), i.e.,  $\hat{\theta}_1$ significant while $\hat{\theta}_2$ nonsignificant), our approach produces undesirably wide CIs: they are four times wider than Wald-type intervals for over 50\%  of Monte Carlo replications. However, this scenario is highly unlikely with a low frequency of only 0.5\%. 

In setting 4,  the results are reported in Table \ref{tab:5}. While both the Wald-type and sequential CIs maintain appropriate marginal coverage probability for $\theta_p=0$, neither method guarantees the conditional coverage level when the trial is early terminated or enters the final stage. In particular, both approaches yield problematic CIs for $\theta_p$, with zero coverage conditioning on $\text{D}_I = \text{S}.$ In a clinical trial context, this implies that researchers might claim benefit for an ineffective treatment, if the group sequential trial is terminated at an early stage by chance. A similar under-coverage issue is observed for the secondary outcome when the primary and secondary outcome estimators are correlated. In contrast, our approach is adaptable to each stage of a group sequential design, providing CIs with correct conditional coverage levels for both primary and secondary outcomes, thereby reducing the risk of misleading conclusions. We also set $\theta_p = 2^{1/2}/10$ and $\theta_s = 2^{1/2}/30$ corresponding to  $\approx 40\%$ power for rejecting $H_0: \theta_p=0$. Similar to the null case, neither the Wald-type nor the sequential CIs maintains correct conditional coverage level when the trial is early terminated or enters the final stage. As a price of ensuring the conditional validity, our approach has a lower power than other two counterparts.

\section{SPRINT data analysis}\label{sec: realData}

\section{SPRINT data analysis}\label{sec: realData}

In this section, we apply our approach to reanalyze the Systolic Blood Pressure Intervention Trial (SPRINT; \citet{sprint2015randomized}) data. SPRINT was a 2-arm, multicenter, randomized clinical trial designed to assess whether an intensive treatment strategy to control systolic blood pressure (SBP) at a lower threshold than the currently recommendation would further reduce cardiovascular disease (CVD) risk. The SPRINT trial enrolled participants aged 50 or older with $\text{SBP} \geq 130$ mm Hg, who were at high risk for CVD. Specifically, SPRINT targeted three high-risk groups of patients: (1) individuals with clinical CVD other than stroke, (2) individuals with chronic kidney disease (estimated glomerular filtration rate [eGFR] 20-59 ml/min/1.73 $\text{m}^2$), and (3) individuals with a high estimated CVD risk based on factors such as SBP, smoking, HDL, LDL, and age. 

The primary outcome considered in SPRINT study is time to a composite endpoint, defined as the first occurrence of any of the following events, i.e., myocardial infarction, acute coronary syndrome, stroke, heart failure, or death from cardiovascular causes, as outlined in Table 2 of \citet{sprint2015randomized}. The times to each individual event also serve as secondary outcomes of the study. We were interested in estimating the hazard ratios for the primary and secondary outcomes between the intensive and standard SBP control groups based on Cox proportional hazards regression. A stopping rule was incorporated based on the test result for the treatment effect on the primary outcome, using a stopping boundary corresponding to O'Brien-Fleming-type $\alpha-$ spending\citep{proschan2006statistical}. As shown in Figure S3 of \citet{sprint2015randomized}, the trial was terminated at the fifth interim analysis after obtaining a promising hazard ratio, with a $z$ score above the stopping boundary of 2.82, corresponding to a one-sided nominal significance level of 0.0024. Subsequently, statistical analysis was conducted to construct $95\%$ CI of the hazard ratios for both the primary and secondary outcomes.

Importantly, the CI of hazard ratio for secondary outcomes should account for the correlation between the primary and secondary outcomes and the conditional nature of early trial termination. To estimate the correlation between these estimators, we employ bootstrapping \citep{tibshirani1993introduction} with 10,000 iterations, and present the result in Table \ref{tab:real}. To illustrate the impact of interim termination on interval estimates, we consider the following conditional statistic and compare the resulting CIs with the conventional Wald-type intervals
\begin{equation}\label{SPRINT_Cons}
    \begin{split}
        & \hat{\theta}_{s}\mid |\hat{\theta}_{p}| > 2.82\hat{\sigma}_{p}n^{-1/2}.
    \end{split}
\end{equation}

Based on results in Table \ref{tab:real}, our approach generally produces wider $95\%$ CIs than the conventional Wald-type counterparts to account for the \textit{selective process} induced by early termination, leading to some discrepancies in conclusions between the two methods. For example, our approach yields borderline acceptance of the null hypothesis that intensive SBP treatment is ineffective in reducing the hazard for the time to the composite endpoint and myocardial infarction, whereas the conventional approach rejects the corresponding null hypotheses with a 95\% confidence.  {Noting that $\hat{\theta}_{p}=-0.31$ and 
$$ \widehat{\Sigma}=\left(\begin{array}{cc}64.46 & 62.01\\ 62.01 & 162.58 \end{array}\right)$$
in SPRINT, where the secondary outcome of interest is the time to myocardial infarction. In this setting, a log-hazard ratio below -0.36, i.e., $\hat{\theta}_{s}\le -0.36,$ yeilds a conditional 95\% CI for $\theta_s$ not containing zero. In other words, the $Z$ score, $n^{1/2}\hat{\theta}_{s}/\hat{\sigma}_{s}$ needs be less than $-2.73$ in order to reject the null $\theta_s=0$ at the two-sided significance level of 0.05. In comparison with the conventional critical value of $-1.96,$ this is a mild price for ensuring the conditional validity. This price varies with the covariance matrix $\widehat{\Sigma}$ and the distance between $\hat{\theta}_{p}$ and the rejection boundary.}  

The condition $|\hat{\theta}_{p}| > 2.82\hat{\sigma}_{p}n^{-1/2}$ in \eqref{SPRINT_Cons} does not fully capture the condition that ``the trial was early terminated at the fifth interim look". Specifically, additional conditions based on the primary outcome estimator and decision boundaries for each of the four preceding interim looks should also be considered. This is similar to the scenario in Tables \ref{tab:5}, where the condition $(\text{D}_I = \text{N},\text{D}_F = \text{S})$ incorporates earlier decision points by setting $\text{D}_I = \text{N}$. Here, due to data limitations, these preceding conditions are not included in our analysis. 

In summary, unlike the significant conclusion drawn from the conventional Wald-type CI, our approach suggests caution in using presented CIs quantifying the treatment effect of the intensive blood pressure control. These new CIs are wider and more likely include the null value.  Despite this limitation, our approach offers CIs with a correct conditional coverage level, making it a valuable supplementary reference for researchers. It provides robust ranges for possible true outcome values, particularly in scenarios where \textit{selective processes} play a significant role in the subsequent usage of those CIs.

\section{Discussion}\label{sec: Dis}
If all CIs were reported and used without depending on observed data from which CIs were constructed, then maintaining conventional marginal coverage probability of relevant CIs is sufficient. However, in practice, a CI is oftentimes reported and interpreted differently depending a \textit{selective process}.  The selection criterion can be implicit and individualized.  When it is explicit, we introduce a novel method for constructing CIs with exact coverage probability, conditional on several specific \textit{selective processes}. These constructed CIs control the FWER, without requiring additional critical value adjustments for multiplicity. Our simulation studies validate the theoretical guarantees for our method and illustrate its adaptability to various clinical scenarios, including post-selection inference and ensuring exact conditional coverage probability at each stage of a group sequential trial. The real data analysis further highlights the potential pitfalls of conventional approaches, such as two-sided Wald tests, which may yield overly optimistic conclusions if \textit{selective processes} are ignored. By accounting for \textit{selective processes}, our approach provides new results for estimating the effects of intensive SBP treatment on the primary outcome and myocardial infarction in the SPRINT.

Despite many merits of our proposal, two main directions warrant future research. First, our approach may yield undesirably wide CIs, as evidenced by the results from simulation. In practice, observing a large $|T_n|$ would increase the chance that $\hat{\theta}_p$ is well separated from its boundary, resulting in a narrower CI.  In a group sequential design, one may monitoring $|T_n|$ in addition to $\hat{\theta}_{p}$ for potential early study termination. For example, a study would be stopped for efficacy, if both $|T_n|$ and $\hat{\theta}_{p}$ were sufficiently large. This new stopping rule would limit the loss in conditional power for testing $\theta_s$ in addition to type I error control for testing $\theta_p.$ 

Second, our approach currently relies on an approximate multivariate normal distribution assumption, which limits its ability in scenarios where outcome estimators are non-Gaussian because of, e.g., small sample size. A critical feature of the multivariate normal distribution leveraged by our approach is the fact that zero-covariance indicate independence between two normally distributed random variables. This property allows us to construct the pivotal quantity under normality by projecting $\hat{\theta}_{p}$ onto a space orthogonal to $\hat{\theta}_{s}.$ However, in cases where those estimators are non-Gaussian, this zero-covariance-to-independence relationship does not hold, complicating the direct application of our approach in such scenarios. Extending our method remains a challenging yet important direction for future research.

\section{Supplementary File}
\section*{S1.   Technical proofs}
\subsection*{S1.1.  Propositions}
\begin{proposition}\label{Supp: prop0}
    Suppose that $(Y_{i1}, Y_{i2}, R_i)_{i=1}^{n}$ are $n$ $i.i.d.$ observations, where $Y_i=(Y_{i1}, Y_{i2})^T$ is a pair of outcomes with a finite 3rd moment and  $R_i$ is a treatment indicator following Bernoulli distribution with $P(R_i=1)=p_0.$ Let $\theta_{p}=E(Y_{i1}|R_i=1)-E(Y_{i1}|R_i=0)$ and $\theta_s=E(Y_{i2}|R_i=1)-E(Y_{i1}|R_i=0)$ are two parameters measuring the average treatment effect on $Y_{i1}$ and $Y_{i2},$ respectively. $\theta_p$ and $\theta_s$ can be estimated by 
    $$\hat{\theta}_{p}=\frac{1}{n_1}\sum_{i=1}^n Y_{i1}R_i -\frac{1}{n_0} \sum_{i=1}^n Y_{i1}(1-R_i) ~\mbox{ and }~ \hat{\theta}_{s}=\frac{1}{n_1}\sum_{i=1}^n Y_{i2}R_i -\frac{1}{n_0} \sum_{i=1}^n Y_{i2}(1-R_i),$$
    respectively, where $n_1=\sum_{i=1}^n R_i$ and $n_0=n-n_1.$  Then condition (C1) is satisfied for $\gamma = \frac{1}{2}.$
\begin{proof}
    When $\hat{\theta}_{p}$ and $\hat{\theta}_{s}$ can be represented as sum of i.i.d elements, and under the finite 3rd moment condition, it is known that
    \begin{equation}\label{BEbound}
        \begin{split}
            & \sup_{(x_p, x_s)\in R^2}\left|F_n(x_p,x_s) - \Phi_{\mu,n^{-1}\Sigma}(x_p,x_s)\right| \leq \frac{C^*}{n^{1/2}}
        \end{split}
    \end{equation}
    by applying Theorem 1.3. of \citep{gotze1991rate} and choosing 
    \begin{equation}\label{biVar}
        \begin{split}
            & S_n \coloneqq \sum_{i=1}^nW_i,~~W_i\coloneqq (n\Sigma)^{-1/2}Y_i,~~f_{x_p,x_s}(t_p,t_s)\coloneqq\mathbbm{1}(t_p\leq x_p, t_s \leq x_s),
        \end{split}
    \end{equation}
    where $Y_i = (Y_{i1},Y_{i2})^T$ is a bi-variate random vector satisfying $E\|Y_i\|^3 < \infty$ (with $\|\cdot\|$ denoting the Euclidean norm) and $\Sigma$ is the covariance matrix of $Y_i$.
\end{proof}
\end{proposition}

\begin{proposition}\label{Supp: prop1}
    Suppose that $(T_{i1}, T_{i2}, C_i, R_i)_{i=1}^{n}$ are $n$ $i.i.d.$ random vectors, where $T_i=(T_{i1}, T_{i2})^T$ is a pair of time to event outcomes with a finite bound, $C_i$ is a bounded censoring time,  and $R_i$ is treatment indicator following  Bernoulli distribution with $P(R_i=1)=p_0.$   The observed data are 
    $(X_{i1}, \delta_{i1}, X_{i2}, \delta_{i2}, R_i)_{i=1}^n,$ where $X_{ij}=\min(T_{ij}, C_{ij})$
    and $\delta_{ij}=I(T_{ij}<C_{ij}), j=1, 2.$ First, we assume that $T_i$ and $C_i$ are independent conditional on $R_i.$  We also assume that $T_{ij}\mid R_i$ follows the proportional hazards assumption:
    $$ h_1(t|R=1)=h_1(t|R=0)\exp(\theta_p) ~~\mbox{and}~~ h_2(t|R=1)=h_2(t|R=0)\exp(\theta_s),$$
    where $h_j(t|R=r)$ is the hazard function of $T_{ij}\mid R_i=r, r\in \{0, 1\}.$
    Let $\hat{\theta}_{p}$ and $\hat{\theta}_{s}$ be the estimated regression coefficient of $R_i$ in respective Cox regression model. Then condition (C1) is satisfied for $\gamma = \frac{1}{3}.$
\begin{proof}
    When $\hat{\theta}_{p}$ and $\hat{\theta}_{s}$ denote the estimates of the treatment effects for $X_{i1}$ and $X_{i2}$ obtained from two Cox models, that is, $\theta_p$ and $\theta_s$. The score function for a Cox model is given by,
\begin{equation}\label{Usurv}
    \begin{split}
        & U_n(\theta_p,X_p) = \sum_{i=1}^n\int_0^{X_p} \left(R_i - \frac{\sum_{l=1}^n R_l\exp\{R_l\theta_p\}Y_{l1}(s)}{\sum_{l=1}^n \exp\{R_l\theta_p\}Y_{l1}(s)}\right)dM_i(s,\theta_p),
    \end{split}
\end{equation}
where $Y_{i1}(s) = \mathbbm{1}(X_{i1} \geq s)$, $N_{i1}(s) = \mathbbm{1}(X_{i1} \leq s, \delta_{i1} = 1)$, $M_i(s,\theta_p)\coloneqq N_{i1}(s) - A_{i1}(s,\theta_p)$, with $A_{i1}(s,\theta_p)$ denoting the compensator process of $N_{i1}(s)$, $X_p = \sup_{s}\{X_{i1}(s) > 0, i=1,\ldots,n \}$ represents the maximal time point when the at-risk set is non-empty. In practice, we can assume that $X_p$ is finite, and therefore, it follows that
\begin{equation}
    \begin{split}
        & \frac{\sum_{l=1}^n R_l\exp\{R_l\theta_p\}Y_{l1}(s)}{\sum_{l=1}^n \exp\{R_l\theta_p\}Y_{l1}(s)} = K(\theta_p,s) + O_p(n^{-1/2}),
    \end{split}
\end{equation}
where $K(\theta_p,s) = E(R_l\exp\{R_l\theta_p\}Y_{l1}(s))/E(\exp\{\theta_p\}Y_{l1}(s))$ represents the risk-weighted covariate and the term $O_p(n^{-1/2})$ is free of $s$ because $X_p$ and $\{R_l\}_{l=1}^n$ are assumed to be finite. This implies that \eqref{Usurv} can be decomposed into the sum of $i.i.d$ random variables and a residual term,
$$U_n(\theta_p,X_p) = \sum_{i=1}^n H_{pi} + \sum_{i=1}^nR_{pi},$$with
\begin{equation}
    \begin{split}
        & H_{pi} = \int_0^{X_p}\left(R_i - K(\theta_p,s)\right)dM_i(s,\theta_p),\\
        & R_{pi} = \int_0^{X_p}O_p(n^{-1/2})dM_i(s,\theta_p).
    \end{split}
\end{equation}
An analogous decomposition can be derived for $U(\theta_s,X_s)$, yielding $H_{si}$ and $R_{si}$, respectively. Following a procedure similar to that in \eqref{biVar}, one can construct $S_n$ based on the bi-variate random vector $(H_{pi}, H_{si})$ with a moderate assumption that $E((H_{pi}^2 + H_{si}^2)^{3/2})$ is bounded. This leads to a Berry-Esseen bound for $(\sum_{i=1}^nH_{pi},\sum_{i=1}^nH_{si})$ as stated in \eqref{BEbound}.

To handle the remainder terms $\sum_{i=1}^n R_{pi}$ and $\sum_{i=1}^n R_{si}$, we can directly extend Lemma 4.1. of \citet{gu1992edgeworth} to the bivariate case.

\begin{lemma}\label{Supp: Lemma}
    Suppose $U_{pn}$, $U_{sn}$ and $V_{pn}$, $V_{sn}$ are four random variables with
    $$\text{pr}(|U_{pn} - V_{pn}| \geq C_pn^{1/6})\leq \frac{C_p'}{n^{1/3}},~~\text{pr}(|U_{sn} - V_{sn}| \geq C_sn^{1/6})\leq \frac{C_s'}{n^{1/3}},$$
    and $U_{pn}$ and $U_{sn}$ satisfy
    $$\sup_{(x_p,x_s)\in R^2}\left|\text{pr}(U_{pn}\leq x_pn^{1/2}, U_{s,n}\leq x_sn^{1/2}) - \Phi_{0,\Sigma}(x_p,x_s)\right|\leq \frac{C_u^*}{n^{1/2}},$$it follows that 
    $$\sup_{(x_p,x_s)\in R^2}\left|\text{pr}(V_{pn}\leq x_pn^{1/2}, V_{sn}\leq x_sn^{1/2}) - \Phi_{0,\Sigma}(x_p,x_s)\right|\leq \frac{C_v^*}{n^{1/3}},$$where $C_p'$, $C_s'$, $C_u^*$ and $C_v^*$ are positive constants, $\Sigma$ represents the covariance between $n^{-1/2}U_{pn}$ and $n^{-1/2}U_{sn}$.
    \begin{proof}
        Let $t_p = x_pn^{1/2}$ and $t_s = x_sn^{1/2}$, we have
        \begin{equation}
            \begin{split}
                & \text{pr}(V_{pn}\leq t_p, V_{sn}\leq t_s)\\
                & = \text{pr}{(V_{pn}\leq t_p, V_{sn}\leq t_s, |U_{pn} - V_{pn}|\leq C_pn^{1/6}, |U_{sn} - V_{sn}|\leq C_sn^{1/6})} + \\
                & \text{pr}(|U_{pn} - V_{pn}| \geq C_pn^{1/6}) + \text{pr}(|U_{sn} - V_{sn}| \geq C_sn^{1/6}),\\
                & \leq \text{pr}(U_p\leq t_p + C_pn^{1/6}, U_s \leq t_s + C_sn^{1/6}) + \frac{2C_s'}{n^{1/3}}.
            \end{split}
        \end{equation}
        Likewise, with $|U_{pn} - V_{pn}|\leq C_pn^{1/6}$ and $V_{pn} > t_p$, we have $U_{pn} > t_p - C_pn^{1/6}$. Therefore, it holds that
        \begin{equation}
            \begin{split}
                & \text{pr}(V_{pn}\leq t_p, V_{sn}\leq t_s)\\
                & \geq \text{pr}(U_p\leq t_p - C_pn^{1/6}, U_s \leq t_s - C_sn^{1/6}) - \frac{2C_s'}{n^{1/3}}.
            \end{split}
        \end{equation}
        In summary, we have
        \begin{equation}
            \begin{split}
                & \sup_{(x_p,x_s)\in R^2}\left|\text{pr}(V_{pn}\leq x_pn^{1/2}, V_{sn}\leq x_sn^{1/2}) - \Phi_{0,\Sigma}(x_p,x_s)\right|\\
                & \leq \sup_{(x_p,x_s)\in R^2} \left|\text{pr}(U_{pn}\leq (x_p \pm \frac{C_p}{n^{1/3}})n^{1/2},U_{sn}\leq (x_s \pm \frac{C_s}{n^{1/3}})n^{1/2})- \Phi_{0,\Sigma}(x_p,x_s)\right| + \frac{2C_s'}{n^{1/3}},\\
                & \leq \frac{2C_s'}{n^{1/3}} + \frac{C_u^*}{n^{1/2}} + \sup_{(x_p,x_s)\in R^2}\left|\Phi_{0,\Sigma}(x_p\pm\frac{C_s}{n^{1/3}}, x_s\pm\frac{C_s}{n^{1/3}}) - \Phi_{0,\Sigma}(x_p,x_s)\right|,\\
                & \leq \frac{2C_s'}{n^{1/3}} + \frac{C_u^*}{n^{1/2}} + \frac{C_{\delta}}{n^{1/3}},
            \end{split}
        \end{equation}
        where the last inequality is because $\Phi_{0,\Sigma}$ has bounded derivative.
    \end{proof}
\end{lemma}

Finally, by choosing $U_{pn} = \sum_{i=1}^n H_{pi}$, $U_{sn} = \sum_{i=1}^n H_{si}$, $V_{pn} = U_n(\theta_p,X_p)$ and $V_{sn} = U_n(\theta_s,X_s)$ and applying Chebyshev's inequality on $\sum_{i=1}^n R_{pi}$, the fact that $M_i(s,\theta_p)$ is a mean-zero martingale and $\text{Var}(R_{pi}) = O(n^{-1})$, it follows that
\begin{equation}
    \begin{split}
        & \text{Pr}\left(\left|\sum_{i=1}^nR_{pi}\right|\geq Cn^{1/6}\right)\leq \frac{C'}{n^{1/3}}.
    \end{split}
\end{equation}
These results, in sum, yield a Berry-Essen type of convergence bound for the distribution of $(\hat{\theta}_{p}, \hat{\theta}_{s})$ toward $\Phi_{\mu,n^{-1}\Sigma}(x_p,x_s)$ at a rate of $n^{-1/3}$.

\end{proof}
\end{proposition}

\subsection*{S1.2.  Main theorem}
To begin with, we restate our conditions,
\begin{enumerate}
    \item[(C1)] The cumulative distribution function (CDF) of $\hat{\mu}$, $F_n(\cdot,\cdot)$, satisfies a uniform convergence toward $\Phi_{\mu,n^{-1}\Sigma}(\cdot,\cdot)$ at a specific rate, namely,
    \begin{equation}\label{CDFConverge}
        \begin{split}
            & \sup_{(x_p, x_s)\in R^2}\left|F_n(x_p,x_s) - \Phi_{\mu,n^{-1}\Sigma}(x_p,x_s)\right| \leq \frac{C^*}{n^{\gamma }},
        \end{split}
    \end{equation}
    where $C^*$ and $\gamma$ are positive constants.
    \item[(C2)] $\hat{\Sigma}_n$ is a weakly consistent estimator of $\Sigma$ and $\min(\sigma_s, \sigma_p)>0.$
    \item[(C3)] The probability of observing $\hat{\theta}_{p} > c_{n}\coloneqq c_p\hat{\sigma}_{p}n^{-1/2}$ is at least $\alpha_p$, with $\alpha_p > 0$.
\end{enumerate}

The following lemma claims if the uniform convergence of the empirical CDF holds, the composition of the empirical CDF with the inverse of the limiting CDF approaches an identity function uniformly over $[0,1]$.

\begin{lemma}\label{Supp: lemma}
    Suppose $F_n(\cdot)$ uniformly converges to $G_n(\cdot)$, i.e.,
    $$\lim_{n\to+\infty}\sup_{x\in\mathbbm{R}}\left|F_n(x) - G_n(x)\right| = 0,$$  where $G_n(x)$ is a continuous CDF and monotonically increasing over $\mathbbm{R}$ for $\forall n$, it follows that 
    $$\lim_{n\to+\infty}\sup_{u\in[0,1]}\left|u - F_n(G_n^{-1}(u))\right| = 0.$$
    \begin{proof}
        We prove the contrapositive. Suppose $\exists~a_0 > 0$, such that we can find a subsequence of $1,\ldots,n$, without loss of generality (WLOG), $1,\ldots,m$, such that,
        $$\sup_{u\in[0,1]}|u - F_m(G_m^{-1}(u))| > a_0.$$ Further choose a subsequence of $1,\ldots,m$, called $1,\ldots,k,$ WLOG, such that
        $$|u_k - F_k(G_k^{-1}(u_k))| > a_0,~~\text{as}~k\to\infty,~\text{for}~u_k\in[0,1],k=1,2,\ldots$$ However, this contradicts $$\lim_{n\to+\infty}\sup_{x\in\mathbbm{R}}\left|F_n(x) - G_n(x)\right| = 0,$$ because 
        $$a_0 < \lim_{k\to+\infty}|u_k - F_k(G_k^{-1}(u_k))| = \lim_{k\to+\infty}|G_k(x_k) - F_k(x_k)| = 0,$$ with $x_k$ denoting $G_k^{-1}(u_k)$.
    \end{proof}
\end{lemma}

Our main theorem can be proved by applying the defined assumptions and Lemma \ref{Supp: lemma}.

\begin{theorem}\label{Supp: mainTheorem}
    Under conditions (C1) - (C3), we have
    \begin{equation}\label{UnifRV}
\lim_{n\rightarrow \infty} \int\text{Pr}\left(G_{\theta_{s},\hat{\sigma}_{s}}^{[a_n^{\delta_n}(t),b_n^{\delta_n}(t)]}(\hat{\theta}_{s})\leq u \biggm |  \hat{\theta}_{p} > c_{n}\right) d H_{T_n\mid \hat{\theta}_{p} > c_{n}}(t) \stackrel{p}{\to} u 
    \end{equation}
 for $\forall u \in [0,1].$ 
In other words, the distribution of $G_{\theta_s,\hat{\sigma}_{s}}^{[a_n^{\delta_n}(T_n),b_n^{\delta_n}(T_n)]}(\hat{\theta}_{s})$ conditioning on $\hat{\theta}_{p} > c_{n}\coloneqq c_p\hat{\sigma}_{p}n^{1/2}$ asymptotically follows $\text{U}(0,1).$ 

\begin{proof}
    The result presented in \eqref{UnifRV} implies the following equality,
    \begin{equation}\label{CDFCond1}
        \begin{split}
            & \int \text{Pr}\left(G_{\theta_s,\hat{\sigma}_{s}}^{[a_n^{\delta_n}(t),b_n^{\delta_n}(t)]}(\hat{\theta}_{s})\leq u\mid \hat{\theta}_{p} > c_{n}\right) d H_{T_n\mid \hat{\theta}_{p} > c_{n}}(t),\\
            & = \int F_{s,n}^{[(a_n^{\delta_n}(t)),b_n^{\delta_n}(t)]}({G_{\theta_s,\hat{\sigma}_{s}}^{[a_n^{\delta_n}(t),b_n^{\delta_n}(t)]}}^{-1}(u)) d H_{T_n\mid \hat{\theta}_{p} > c_{n}}(t),
        \end{split}
    \end{equation}
    where $F_{s,n}^{[(a_n^{\delta_n}(t)),b_n^{\delta_n}(t)]}(\cdot)$ denotes the CDF of $F_n(+\infty,\cdot)$ truncated between $(a_n^{\delta_n}(t))$ and $b_n^{\delta_n}(t)$. The remainder is to evaluate the difference between the right hand side of \eqref{CDFCond1} and $u$, of which the absolute difference after limitation is given by
    \begin{equation}\label{CDFeq1}
        \begin{split}
            & \left|\lim_{n\to +\infty}\int \left(F_{s,n}^{[(a_n^{\delta_n}(t)),b_n^{\delta_n}(t)]}({G_{\theta_s,\hat{\sigma}_{s}}^{[a_n^{\delta_n}(t),b_n^{\delta_n}(t)]}}^{-1}(u)) - u\right) d H_{T_n\mid \hat{\theta}_{p} > c_{n}}(t)\right|\\
            &  \leq \lim_{n\to +\infty}\int \left(\mathbbm{1}(|t| \leq C_n) + \mathbbm{1}(|t| > C_n)\right) \left|F_{s,n}^{[(a_n^{\delta_n}(t)),b_n^{\delta_n}(t)]}({G_{\theta_s,\hat{\sigma}_{s}}^{[a_n^{\delta_n}(t),b_n^{\delta_n}(t)]}}^{-1}(u)) - u\right| \\
            & d H_{T_n\mid \hat{\theta}_{p} > c_{n}}(t),
        \end{split}
    \end{equation}
    where $C_n$ is a sequence equal to $c'(2\sigma_s^2\log(n))^{1/2}n^{-1/2}$, with $c'$ being a positive value defined later. The last display of inequality (c) can be further upper bounded by
    \begin{equation}\label{CDFeq2}
        \begin{split}
            & (1). \lim_{n\to +\infty} \int_{|t| \leq C_n}\left|\mathbbm{1}(|\hat{\sigma}_{ps}| > \delta_n,\hat{\sigma}_{s}^2 >\delta_n)(F_{s,n}^{[(a_n^{\delta_n}(t)),b_n^{\delta_n}(t)]}({G_{\theta_s,\hat{\sigma}_{s}}^{[a_n^{\delta_n}(t),b_n^{\delta_n}(t)]}}^{-1}(u)) - u)\right|  \\
            & d H_{T_n\mid \hat{\theta}_{p} > c_{n}}(t) + \\
            & (2).  \lim_{n\to +\infty} \int_{|t| \leq C_n}\left|\mathbbm{1}(|\hat{\sigma}_{ps}| \leq \delta_n,\hat{\sigma}_{s}^2 >\delta_n)(F_{s,n}^{[(a_n^{\delta_n}(t)),b_n^{\delta_n}(t)]}({G_{\theta_s,\hat{\sigma}_{s}}^{[a_n^{\delta_n}(t),b_n^{\delta_n}(t)]}}^{-1}(u)) - u)\right|  \\
            & d H_{T_n\mid \hat{\theta}_{p} > c_{n}}(t) + \\
            & (3). \lim_{n\to +\infty} \int_{|t| \leq C_n}\left|\mathbbm{1}(\hat{\sigma}_{s}^2 \leq\delta_n)(F_{s,n}^{[(a_n^{\delta_n}(t)),b_n^{\delta_n}(t)]}({G_{\theta_s,\hat{\sigma}_{s}}^{[a_n^{\delta_n}(t),b_n^{\delta_n}(t)]}}^{-1}(u)) - u)\right|  \\
            & d H_{T_n\mid \hat{\theta}_{p} > c_{n}}(t) + \\
            & (4). \lim_{n\to +\infty} \int_{|t| > C_n}\left|F_{s,n}^{[(a_n^{\delta_n}(t)),b_n^{\delta_n}(t)]}({G_{\theta_s,\hat{\sigma}_{s}}^{[a_n^{\delta_n}(t),b_n^{\delta_n}(t)]}}^{-1}(u)) - u\right|d H_{T_n\mid \hat{\theta}_{p} > c_{n}}(t).
        \end{split}
    \end{equation}
    We proceed by upper bounding \eqref{CDFeq2}. Firstly, by choosing a $\delta_n$ sufficiently small, or for instance, $O(n^{-1/2}(\log(n))^{1/2})$, Part 3 converges to 0 in probability 1 by Assumption (C2) that $\hat{\sigma}_{s}^2$ is a weakly consistent estimator of $\sigma_s$, which is greater than 0. Part 2 can be upper bounded by Lemma \ref{Supp: lemma} since it holds that $(a_n^{\delta_n}(t),b_n^{\delta_n}(t)) = (-\infty,+\infty)$ following their definition and Assumption (C1) which implies the point-wise convergence and hence uniform convergence since $\Phi_{\mu,n^{-1}\hat{\Sigma}}(\cdot,\cdot)$ is continuous. Part 1 requires detailed clarifications. For simplicity, we define
    \begin{equation}\label{eqSimp}
        \begin{split}
            & \Delta F(u,v) = F_{s,n}(v) - F_{s,n}(u),\\
            & \Delta G_{\theta_s,\hat{\sigma}_{s}}(u,v) = G_{\theta_s,\hat{\sigma}_{s}}(v) - G_{\theta_s,\hat{\sigma}_{s}}(u),
        \end{split}
    \end{equation}
    where $F_{s,n}(\cdot)$ is equal to $F_n(+\infty,\cdot)$. The absolute difference between the two truncated CDFs $\frac{\Delta F(a_n^{\delta_n}(t),u)}{\Delta F(a_n^{\delta_n}(t),b_n^{\delta_n}(t))}$ and $\frac{\Delta G_{\theta_s,\hat{\sigma}_{s}}(a_n^{\delta_n}(t),u)}{\Delta G_{\theta_s,\hat{\sigma}_{s}}(a_n^{\delta_n}(t),b_n^{\delta_n}(t))}$ is given by
    \begin{equation}\label{eqDiff}
        \begin{split}
            & \left|\frac{\Delta F(a_n^{\delta_n}(t),u)}{\Delta F(a_n^{\delta_n}(t),b_n^{\delta_n}(t))} - \frac{\Delta G_{\theta_s,\hat{\sigma}_{s}}(a_n^{\delta_n}(t),u)}{\Delta G_{\theta_s,\hat{\sigma}_{s}}(a_n^{\delta_n}(t),b_n^{\delta_n}(t))}\right|,\\
            & \leq \frac{1}{|\Delta F(a_n^{\delta_n}(t),b_n^{\delta_n}(t))|}\left|\Delta F(a_n^{\delta_n}(t),u) - \Delta G_{\theta_s,\hat{\sigma}_{s}}(a_n^{\delta_n}(t),u)\right| + ,\\
            & \frac{1}{|\Delta F(a_n^{\delta_n}(t),b_n^{\delta_n}(t))|}\left|\Delta F(a_n^{\delta_n}(t),b_n^{\delta_n}(t)) - \Delta G_{\theta_s,\hat{\sigma}_{s}}(a_n^{\delta_n}(t),b_n^{\delta_n}(t))\right|,\\
            & \leq \frac{2C^*}{|\Delta F(a_n^{\delta_n}(t),b_n^{\delta_n}(t))|}n^{-\gamma},
        \end{split}
    \end{equation}
    where the last inequality is given by applying Assumption (C1). Note that $|\Delta F(a_n^{\delta_n}(t),b_n^{\delta_n}(t))|$ takes its lower bound by taking $t$ at its boundary, yielding a lower bound proportional to $n^{-{c'}^2}$ by using the Chernoff's bound together with Assumption (C1). By taking ${c'}^2 < \gamma$ guarantees the uniform convergence of $\frac{\Delta F(a_n^{\delta_n}(t),u)}{\Delta F(a_n^{\delta_n}(t),b_n^{\delta_n}(t))}$ to $\frac{\Delta G_{\theta_s,\hat{\sigma}_{s}}(a_n^{\delta_n}(t),u)}{\Delta G_{\theta_s,\hat{\sigma}_{s}}(a_n^{\delta_n}(t),b_n^{\delta_n}(t))}$ since the latter CDF is continuous. This further upper bounds Part 3 again by applying Lemma \ref{Supp: lemma}. To upper bound Part 4, we proceed by
    \begin{equation}\label{eqLast}
        \begin{split}
            & \int_{|t| > C_n}\left|F_{s,n}^{[(a_n^{\delta_n}(t)),b_n^{\delta_n}(t)]}({G_{\theta_s,\hat{\sigma}_{s}}^{[a_n^{\delta_n}(t),b_n^{\delta_n}(t)]}}^{-1}(u)) - u\right|d H_{T_n\mid \hat{\theta}_{p} > c_{n}}(t),\\
            & \leq 2\int_{|t| > C_n}d H_{T_n\mid \hat{\theta}_{p} > c_{n}}(t),\\
            & \leq 2\alpha_p\text{Pr}(\left|T_n\right| > C_n).
        \end{split}
    \end{equation}
    Note that by Assumption (C3), the variance of $n^{1/2}T_n$ converges to $(1 - \frac{\sigma_{ps}}{\sigma_{p}\sigma_{s}})\sigma_p^2$, which is a positive value. Then by the definition of $C_n$ and the Chernoff's bound, the last display of \eqref{eqLast} can be upper bounded by a value that converges to 0.
\end{proof}
\end{theorem}

\begin{table}[h]
    \centering
        \caption{Comparison of the proposed method and conventional Wald-type interval across 1,000 Monte Carlo replications for setting 1. COV: empirical coverage level of constructed 95\% CI; Width: median and inter-quartile ranges of the width of constructed 95\% CIs. }
    \label{tab:1}
   \begin{tabular}{lllllllll}
    \toprule
 
 $\rho$ & $n$ &  \multicolumn{2}{c}{Wald-type CI} && \multicolumn{4}{c}{Proposed CI}\\
 & & COV & Width && $\delta_n$ & COV & Width & $\hat{\ell}_{1}$ \\
  \cline{3-4} \cline{6-9}
  \addlinespace[5pt] 
 0.0$^+$  & 100 & 0.957 & 0.537 (0.094) && 0.00 & 0.952 & 0.587 (0.213) & 0.008 \\
     &     &  &  && 0.05 & 0.952 & 0.582 (0.205) & 0.007 \\
     &     &  &  && 0.10 & 0.954 & 0.575 (0.183) & 0.007\\
  & 400 & 0.961 & 0.272 (0.027) && 0.00 & 0.957 & 0.279 (0.040) & 0.011 \\
     &     &  &  && 0.05 & 0.958 & 0.278 (0.038) & 0.011\\
     &     &  &  && 0.10 & 0.957 & 0.275 (0.033) & 0.012\\
  & 1000 & 0.958 & 0.173 (0.011) && 0.00 & 0.957 & 0.175 (0.014) & 0.022 \\
     &     &  &  && 0.05 & 0.956 & 0.174 (0.013) & 0.024\\
     &     &  &  && 0.10 & 0.956 & 0.174 (0.012) & 0.022\\
 0.4$^*$ & 100 & 0.867 & 0.538 (0.084) && 0.00 & 0.948 & 0.853 (0.546) & 0.017\\
     &     &  &  && 0.05 & 0.948 & 0.853 (0.546) & 0.017\\
     &     &  &  && 0.10 & 0.948 & 0.853 (0.546) & 0.017\\
  & 400 & 0.850 & 0.272 (0.024) && 0.00 & 0.951 & 0.445 (0.215) & 0.012 \\
     &     &  &  && 0.05 & 0.951 & 0.445 (0.215) & 0.012\\
     &     &  &  && 0.10 & 0.951 & 0.445 (0.215) & 0.012\\
 & 1000 & 0.880 & 0.174 (0.010) && 0.00 & 0.967 & 0.290 (0.126) & 0.016 \\
    &     &  &  && 0.05 & 0.967 & 0.290 (0.126) & 0.016\\
    &     &  &  && 0.10 & 0.967 & 0.290 (0.126) & 0.016\\ 
  \bottomrule
  \multicolumn{9}{l}{$^+: \sigma_{ps}=0;$ $^*: \sigma_{ps}=0.81 >\delta_n$}
    \end{tabular}

\end{table}

\begin{table}[h]
    \centering
        \caption{Comparison of the proposed method and conventional Wald-type interval across 1,000 Monte Carlo replications for setting 2. COV: empirical coverage level of constructed 95\% CI; Width: median and inter-quartile ranges of the width of constructed 95\% CIs.}
    \label{tab:3}
   \begin{tabular}{lllllllll}
    \toprule
 $\rho$  & $n$ &  \multicolumn{2}{c}{Wald-type CI} && \multicolumn{4}{c}{Proposed CI}\\
 & & COV & Width && $\delta_n$ & COV & Width & $\hat{\ell}_{1}$ \\
  \cline{3-4} \cline{6-9}
  \addlinespace[5pt] 
 0.6$^+$ & 100 & 0.721 & 0.799 (0.013) && 0.00 & 0.945 & 1.188 (1.413) & 0.018\\
    &     &  &  && 0.05 & 0.945 & 1.188 (1.413) & 0.018\\
     &     &  &  && 0.10 & 0.944 & 1.194 (1.249) & 0.020\\
  & 400 & 0.706 & 0.394 (0.002) && 0.00 & 0.949 & 0.571 (0.502) & 0.009 \\
     &     &  &  && 0.05 & 0.949 & 0.571 (0.502) & 0.009\\
    &     &  &  && 0.10 & 0.959 & 0.573 (0.487) & 0.015\\
  & 1000 & 0.697 & 0.249 (0.001) && 0.00 & 0.944 & 0.361 (0.329) & 0.012 \\
     &     &  &  && 0.05 & 0.948 & 0.363 (0.325) & 0.009\\
    &     &  &  && 0.10 & 0.949 & 0.361 (0.335) & 0.011\\
  \bottomrule
  \multicolumn{9}{l}{$^+: \sigma_{ps}=2.66 > \delta_n$ }
    \end{tabular}
\end{table}
\clearpage

{\small
\begin{table}[h]
    \centering
        \caption{Comparison of Conditional CI and Conventional Wald-type Interval across 100,000 Monte Carlo replications for setting 3. COV: empirical coverage level of constructed 95\% CI; Width: median and inter-quartile ranges of the width of constructed 95\% CIs; Power: the empirical probability that CI doesn't include zero under alternative; the Ratio: the median and inter-quartile range of the width ratio of conditional CI vs Wald-type CI}
    \label{tab:4}
   \begin{tabular}{ccccccccccc}
    \toprule
 \multicolumn{3}{c}{} & \multicolumn{3}{c}{Conditional CI} && \multicolumn{3}{c}{Wald-type CI} &  \\
\cline{4-6} \cline{8-10}
\addlinespace[5pt] 
& $\text{D}_1/\text{D}_2$ & Freq & COV & Width & Power && COV & Width & Power & Ratio\\
\multicolumn{11}{c}{Null}\\
$\theta_1$ & $\emptyset, \emptyset$ & 1.000 & 0.949 & 0.534 (0.093) &  && 0.949 & 0.534 (0.093) & & 1.000 (0.000)\\
$\theta_2$ & N, $\emptyset$ & 0.949 & 0.948 & 0.586 (0.141) &  && 0.955 & 0.539 (0.083) & & 1.043 (0.143)\\
  & S, $\emptyset$ & 0.051 & 0.948 & 0.842 (0.492) &  && 0.812 & 0.533 (0.083) & & 1.553 (0.828)\\
    & $\emptyset$, $\emptyset$ & 1.000 & 0.948 & 0.591 (0.157) & && 0.948 & 0.539 (0.083) & & 1.048 (0.177) \\
$\theta_3$ & N, N & 0.906 & 0.950 & 0.627 (0.200) &  && 0.961 & 0.541 (0.080) & & 1.114 (0.297) \\
  & S, N & 0.042 & 0.946 & 1.440 (1.974) &  && 0.855 & 0.535 (0.078) && 2.648 (3.573) \\
  & N, S & 0.043 & 0.951 & 1.456 (1.920) &  && 0.845 & 0.537 (0.078) && 2.630 (3.431)\\
  & S, S & 0.010 & 0.953 & 0.947 (0.809) &  && 0.657 & 0.537 (0.076) && 1.726 (1.540)\\
  & $\emptyset, \emptyset$ & 1.000 & 0.950 & 0.643 (0.264) & && 0.949 & 0.540 (0.080) &&  1.138 (0.429)\\
\multicolumn{11}{c}{Alternative}\\
$\theta_1$ & $\emptyset, \emptyset$ & 1.000 & 0.949 & 0.534 (0.093) & 0.852 && 0.949 & 0.534 (0.093) & 0.852& 1.000 (0.000) \\
$\theta_2$ & N, $\emptyset$ & 0.148 & 0.955 & 0.944 (0.973) & 0.081 && 0.908 & 0.553 (0.089) &0.032& 1.670 (1.749)\\
  & S, $\emptyset$ & 0.852 & 0.947 & 0.584 (0.168) & 0.167 && 0.954 & 0.537 (0.082) &0.206& 1.032 (0.217)\\
    & $\emptyset$, $\emptyset$ & 1.000 & 0.948 & 0.603 (0.219) & 0.154 && 0.947 & 0.539 (0.083) &0.180& 1.059 (0.350)\\
$\theta_3$ & N, N & 0.143 & 0.955 & 1.006 (1.017) & 0.079 && 0.910 & 0.554 (0.085) &0.030& 1.769 (1.829) \\
  & S, N & 0.677 & 0.950 & 0.742 (0.511) & 0.138 && 0.967 & 0.539 (0.079) &0.149& 1.343 (0.905)\\
  & N, S & 0.005 & 0.953 & 2.301 (3.583) & 0.051 && 0.895 & 0.551 (0.080) &0.133& 4.305 (6.951)\\
  & S, S & 0.176 & 0.949 & 0.805 (0.344) & 0.134 && 0.911 & 0.536 (0.078) &0.413& 1.491 (0.634)\\
  & $\emptyset, \emptyset$ & 1.000 & 0.950 & 0.643 (0.264) & 0.128 && 0.949 & 0.540 (0.080) &0.178& 1.138 (0.429)\\
  \bottomrule
    \end{tabular}
\end{table}
}

\begin{table}[h]
    \centering
        \caption{Comparison of Conditional CI, Wald-type CI and Sequential CI across 100,000 Monte Carlo replications for setting 4.  COV: empirical coverage level of constructed 95\% CI; LBV: median of the lower end of constructed one-sided 95\% CIs; Power: the empirical probability that CI doesn't include zero under alternative; }
    \label{tab:5}
   \begin{tabular}{cccc crc c crc c lrl}
    \toprule
 \multicolumn{3}{c}{} && \multicolumn{3}{c}{Conditional CI} && \multicolumn{3}{c}{Wald-type CI} &&  \multicolumn{3}{c}{Sequential CI} \\
 \cline{5-7} \cline{9-11} \cline{13-15}
 \addlinespace[5pt] 
& $\text{D}_I/\text{D}_F$ & Freq && COV & LBV & Power && COV & LBV & Power && COV & LBV & Power\\
\multicolumn{15}{c}{Null}\\
$\theta_p$ & S  $\emptyset$ & 0.01 && 0.955 & -1.192 &&& 0.000 & 0.034 &&& 0.000 & 0.132 \\
    & N  $\emptyset$ & 0.99 && 0.950 & -0.172 &&& 0.958 & -0.167 &&& 0.955 & -0.160 \\
 & N  S & 0.04 && 0.951 & -0.853  &&& 0.000 & 0.028  &&& 0.000 & 0.027  \\
  & N  N & 0.95 && 0.950 & -0.168  &&& 1.000 & -0.172  &&& 0.997 & -0.166  \\
    & $\emptyset$  $\emptyset$ & 1.00 && 0.950 & -0.172  &&& 0.950 & -0.165 &&& 0.947 & -0.159\\
$\theta_s$ & S  $\emptyset$ & 0.01 && 0.937 & -0.564&&& 0.871 & -0.135 &&& 0.599 & -0.023 \\
    & N  $\emptyset$ & 0.99 && 0.950 & -0.169 &&& 0.959 & -0.167  &&& 0.954 & -0.164  \\
 & N  S & 0.04 && 0.954 & -0.457  &&& 0.780 & -0.066  &&& 0.762 & -0.062  \\
  & N  N & 0.95 && 0.950 & -0.166  &&& 0.967 & -0.171  &&& 0.962 & -0.168  \\
    & $\emptyset$  $\emptyset$ & 1.00 && 0.950 & -0.170  &&& 0.958 & -0.167  &&& 0.951 & -0.163 \\
\multicolumn{15}{c}{Alternative}\\
$\theta_p$ & S  $\emptyset$ & 0.10 && 0.948 & -0.713  &0.130&& 0.907 & 0.050  &1.000&& 0.459 & 0.147  & 1.000 \\
    & N  $\emptyset$ & 0.90 && 0.951 & -0.095 &0.214&& 0.977 & -0.034  &0.356&& 1.000 & -0.030  & 0.092 \\
 & N  S & 0.32 && 0.949 & -0.357  &0.173&& 0.936 & 0.050  &1.000&& 1.000 & 0.045  & 0.093 \\
  & N  N & 0.58 && 0.952 & -0.067  &0.236&& 1.000 & -0.079  &0.000&& 1.000 & -0.073 & 0.092 \\
    & $\emptyset$  $\emptyset$ & 1.00 && 0.950 & -0.105  &0.206&& 0.970 & -0.022 &0.418&& 0.948 & -0.018 & 0.179 \\
$\theta_s$ & S  $\emptyset$ & 0.10 && 0.947 & -0.409  &0.086&& 0.951 & -0.157 &0.100&& &&\\
    & N  $\emptyset$ & 0.90 && 0.955 & -0.159  &0.092&& 0.965 & -0.126 &0.095&& &&\\
 & N  S & 0.32 && 0.963 & -0.247  &0.081&& 0.923 & -0.078  &0.188&& &&\\
  & N  N & 0.58 && 0.951 & -0.139 &0.097&& 0.988 & -0.152  &0.044&& &&\\
    & $\emptyset$  $\emptyset$ & 1.00 && 0.954 & -0.167  &0.091&& 0.964 &-0.128& 0.096 &&&&\\
\bottomrule
    \end{tabular}
\end{table}

\begin{table}[h]
    \centering
        \caption{Comparison of $95\%$ CIs using our approach versus the conventional method. Correlations between primary and secondary outcome estimators are estimated from 10,000 bootstrapping iterations. COR: Pearson correlation coefficient calculated between the primary and secondary outcome estimates, with values expressed on the log‐hazard ratio scale. EST: Hazard ratio point estimate.}
    \label{tab:real}
   \begin{tabular}{lllccccc}
    \toprule
\textbf{Primary} & COR && \multicolumn{2}{c}{Wald-type} && Proposed CI  \\
 &  && EST & CI &&  \\
\cline{4-5} \cline{7-7}
\addlinespace[5pt] 
Composite endpoint & 1.000 && 0.737 & (0.626, 0.863) && (0.628, 1.030)  \\ 
    \midrule
\textbf{Secondary} & \multicolumn{6}{c}{}  \\
    \addlinespace[5pt] 
Myocardial infarction & 0.612 && 0.721 & (0.555, 0.931) && (0.556, 1.039)  \\
Acute coronary syndrome & 0.366 && 1.021 & (0.655, 1.584) && (0.655, 1.649)  \\
Stroke & 0.472 && 0.890 & (0.640, 1.230) && (0.639, 1.316)  \\
Heart failure & 0.500 && 0.646 & (0.471, 0.880) && (0.471, 0.953)  \\
Death from CVD & 0.394 && 0.574 & (0.386, 0.843) && (0.386, 0.887) \\
  \bottomrule
    \end{tabular}
\end{table}

\bibliographystyle{chicago}
\bibliography{ref.bib}

\end{document}